# Demonstration of Probabilistic Constellation Shaping for Continuous Variable Quantum Key Distribution

**François Roumestan[(1)], Amirhossein Ghazisaeidi[(1)], Jérémie Renaudier[(1)], Patrick Brindel[(1)], Eleni Diamanti[(2)], and Philippe Grangier[(3)]**

*(1) Nokia Bell Labs, Paris-Saclay, route de Villejust, F-91620 Nozay, France*
*(2) Sorbonne Université, CNRS, LIP6, 4 place Jussieu, F-75005 Paris, France*
*(3) Université Paris-Saclay, IOGS, CNRS, Laboratoire Charles Fabry, 2 avenue Fresnel, F-91127 Palaiseau, France*
*francois.roumestan@nokia.com*

**Abstract:** We demonstrate, for the first time to our knowledge, continuous-variable quantum key distribution using probabilistically-shaped 1024-QAM and true local oscillator, achieving 38.3Mb/s secret key rate over 9.5km, averaged over the transmission time of 100 blocks.

## 1. Introduction

Continuous variable quantum key distribution (CV-QKD), based on Gaussian modulation, has emerged as an attractive technology to implement practical secure communications over insecure physical links, thanks to its compatibility with commercially available light sources and equipment, and has enjoyed a rekindled interest in recent years [1,2]. Several experimental demonstrations have been performed, where either bright pulses are optically time-multiplexed together with the weak quantum signal [3], or frequency domain pilot tones are sent together with the signal [4]. Many experiments use low-cardinality constellations, like QPSK, where security proofs are less advanced than those for Gaussian modulation, and where the number of secret bits per symbol is intrinsically limited [5,6,7]. In this contribution, we demonstrate, for the first time to our knowledge, the possibility of using dual-polarization probabilistically shaped 1024 QAM signal, as an acceptable approximation to Gaussian modulation, with characterized bounded error. We use Nyquist pulse-shaping, 50% QPSK symbols as pilots, digitally multiplexed with the quantum signal, true local oscillator, and standard DSP, to recover the phase and polarization axis, without any extra analog or digital complexity. The measured parameters indicate the feasibility of generating average secret key rates of 38.3 Mbit/s (averaged over 100 blocks of $1.8 \times 10^6$ symbols) through 9.5 km of standard single mode fiber.

## 2. System description

The experimental setup is outlined in Fig. 1a. Alice uses a tunable laser source with 10 kHz nominal full width at half maximum tuned to 1550 nm. She prepares random coherent states using a standard dual polarization I/Q optical modulator controlled by four outputs of a Tektronix 5 GS/s arbitrary waveform generator (AWG). The AWG generates a dual polarization 400 MBaud signal with a digital root-raised cosine (RRC) pulse shape with adjustable roll-off factor $\gamma$. Details on signal modulation and pulse shaping are given in sections 2.1 and 2.2. The coherent states are sent through either a variable optical attenuator (VOA) or a 9.5 km conventional single-mode fiber link with 2.2 dB characterized channel loss. Bob uses a standard integrated coherent receiver (ICR) with a free-running laser source identical to that of Alice as local oscillator. The received signal is sampled using a Teledyne Lecroy 1 GHz real-time oscilloscope with 5 GS/s sampling rate. The sampled waveforms are stored for offline digital signal processing (DSP). The DSP procedure is described in section 2.3. A fast optical switch with 280 ns response time is used to switch off the signal light in order to perform periodic calibration of the shot noise, alternated with signal recording. A microprocessor controls the optical switch and triggers the oscilloscope such that shot noise calibration and signal acquisition are performed consecutively with 0.2 ms delay. Fig. 1b. shows a back to back characterization of the setup where we compare the ideal SNR, computed by calibrating the noise power at Bob's input, and the actual SNR computed after applying the DSP. This allows us to visualize that the DSP exhibits very low impairments, even at SNR values around 0 dB.

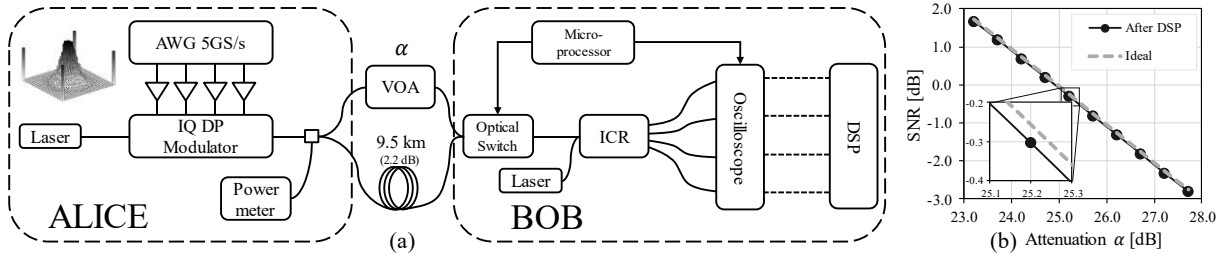

Fig. 1. (a) Experimental testbed featuring 10 kHz linewidth lasers sources, conventional I/Q dual polarization optical modulator, 5GS/s arbitrary waveform generator, conventional integrated coherent receiver, 1 GHz oscilloscope with 5 GS/s sampling rate. (b) Back to back characterization of the setup.

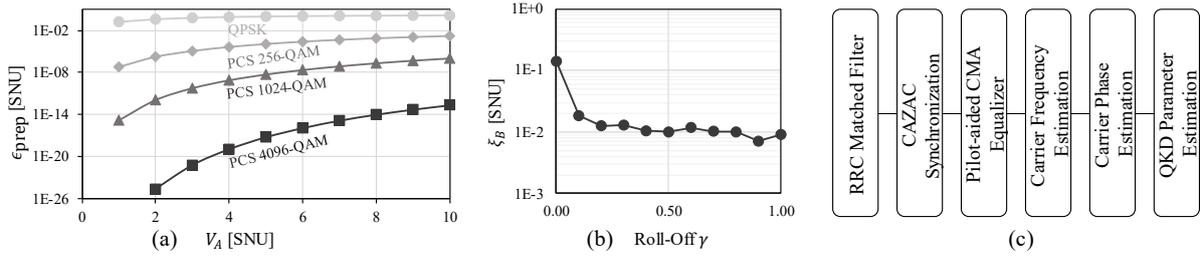

Fig. 2. (a) $\epsilon_{\text{prep}}$ as a function of Alice variance $V_A$ in shot noise unit (SNU) (see text). (b) Measured average excess noise variance at Bob's site vs. roll-off factor $\gamma$ of the RRC filter. (c) Standard DSP procedure.

*2.1. Probabilistic Constellation Shaping with High Cardinality*

The I/Q symbols on both polarizations are chosen randomly according to a probabilistic constellation shaping (PCS) QAM modulation with high cardinality that approximates the Gaussian modulation [8]. The constellation points $X = P + iQ$ of the PCS $2^M$-QAM are the constellation points of the standard $2^M$-QAM. The probability of $X = p + iq$ follows a discrete Boltzmann-Maxwell distribution with free parameter $\nu \geq 0$:

$$P_X(p+iq) = \frac{e^{-\nu(p^2+q^2)}}{\sum_{p,q} e^{-\nu(p^2+q^2)}}.$$

Following the approach in [9], we compute the "preparation error" of the modulation, $\epsilon_{\text{prep}}$, as the trace distance between the quantum states with PCS $2^M$-QAM modulation and ideal Gaussian modulation. The CV-QKD protocol with PCS $2^M$-QAM modulation using Gaussian modulation security proof, assuming coherent attacks is then considered as $(\epsilon + \epsilon_{\text{prep}})$-secure, where $\epsilon$ takes into account other security parameters, related e.g. to finite size (see section 3). In Fig. 2a $\epsilon_{\text{prep}}$ is plotted as a function of the modulation variance of Alice, $V_A$, for QPSK, PCS 256-QAM, PCS 1024-QAM and PCS 4096-QAM modulations, where the parameter $\nu$ is optimized to minimize $\epsilon_{\text{prep}}$ for each point. We observe that PCS 1024-QAM modulation exhibits $\epsilon_{\text{prep}}$ values below $10^{-8}$ with $V_A$ around 5 shot noise units (SNU). Hence, it is suitable for secure key exchange using Gaussian security proofs [9]. The current setup uses PCS 1024-QAM, but we are working towards integrating PCS 4096-QAM in our DSP as it exhibits much lower $\epsilon_{\text{prep}}$, below $10^{-16}$ for $V_A$ around 5 SNU.

*2.2. Pulse Shaping*

The signal is prepared with a root raised cosine (RRC) pulse shape with roll-off factor $\gamma$. Fig. 2b. shows the average excess noise variance at Bob's side as a function of the roll off factor $0 \leq \gamma \leq 1$. The excess noise is the noise that cannot be attributed to the shot noise or the receiver's electrical noise. Whatever its origin, it is therefore attributed to the action of an eavesdropper. We are interested in the lowest excess noise variance possible. We observe that the excess noise variance decreases significantly between $\gamma = 0$ and $\gamma = 0.1$, and then decreases more slowly. We typically use roll-off factors between 0.4 and 1.

In order to avoid low frequency noise, observed on Alice's power drivers and on Bob's ICR power supply, we digitally upconvert the signal by a 500 MHz frequency shift away from DC; the equivalent signal bandwidth is thus extended from 100 to 900 MHz.

*2.3. Digital Signal Processing*

The CV-QKD signal is interleaved in time with a deterministic sequence of QPSK symbols with a 13 dB higher amplitude. This QPSK sequence is public information shared by Alice and Bob. It is used by the DSP to recover the random PCS QAM symbols sent by Alice. The DSP procedure is outlined in Fig. 2c. The sampled outputs from the oscilloscope are assembled as two complex signals, one for each orthogonal polarization. A matched RRC filter is applied to both signals. CAZAC sequences (constant amplitude zero autocorrelation) inserted at the beginning of the pilot sequence are used to synchronize the received signal. Polarization demultiplexing is performed with a constant modulus algorithm (CMA) equalizer updated on the pilots and followed by down-sampling to 1 sample per symbol [10]. Carrier frequency estimation is performed using a periodogram. Pilot-aided maximum-likelihood estimation applied on the pilots and combined with linear interpolation is used for carrier phase estimation. Finally, the QKD parameters are estimated according to the following equation:

$$V_B = \frac{\eta T}{2} V_A + 1 + V_{el} + \xi_B$$

where $V_B$ is Bob's modulation variance, $V_A$ Alice's modulation variance (both: per quadrature per polarization), $\eta$ the quantum efficiency of the receiver, $T$ the channel transmittance, $V_{el}$ the receiver's electrical noise variance, $\xi_B$ the excess noise at Bob's side and 1 stands for the shot noise variance. The excess noise at Alice's side is simply $\xi_A = 2\xi_B/(\eta T)$.

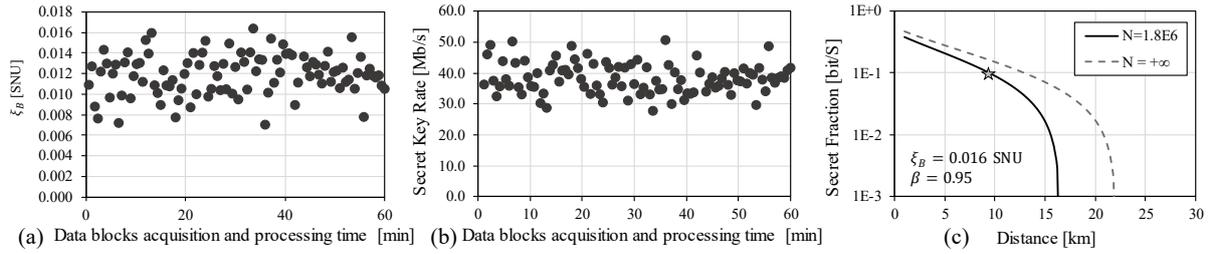

Fig. 3. (a) Excess noise variance at Bob's site $\xi_B$ vs. time over 9.5 km SMF (2.2 dB loss) with PCS 1024-QAM, $V_A = 5$ SNU, $\gamma = 1$, and $N = 1.8 \times 10^6$ symbols per marker point. (b) Secret key rate vs. time for the same parameters, computed with worst case estimator and reconciliation efficiency $\beta = 0.95$.
(c) Secret fraction vs. distance, with $\xi_B = 0.016$ SNU and $\beta = 0.95$. The finite size worst case estimator is 0.025 SNU with security parameter $\epsilon = 10^{-8}$. (SNU: shot noise unit.)

### 3. Results

Fig. 3a. shows the measured excess noise variance $\xi_B$ at Bob's side in SNU for 100 blocks over 1 hour of experiment, with the 9.5 km fiber link. The average $\xi_B$ is 0.012 SNU while the maximal value is 0.016 SNU. The excess noise variance is estimated using the standard estimator of the variance with $N = 1.8 \times 10^6$ symbols. Fig. 3b. shows the secret key rate for the corresponding excess noise of Fig. 3a, computed with the worst case estimator for $N = 1.8 \times 10^6$ symbols and $\epsilon$-security with $\epsilon = 10^{-8}$, and a reconciliation efficiency $\beta = 0.95$. The block-averaged secret key rate over 100 blocks is 38.3 Mb/s with minimal and maximal values 27.7 and 50.7 Mb/s.

The worst case estimator for the maximal measured $\xi_B = 0.016$ is 0.025 SNU. Fig. 3c. shows the secret fraction in bits per symbol as a function of distance for the worst case estimator (solid line) and without considering finite size effects (asymptotic rate, dashed line). The star shows the actual measured distance. Distances up to 16 km can be achieved with the present state of the experimental testbed.

### 4. Conclusion

We investigated the use of PCS QAM modulations for CV-QKD. We established that PCS 1024 QAM and higher cardinality PCS QAM modulations are close enough approximations of the Gaussian modulation to ensure security corresponding to $\epsilon_{\text{prep}} \leq 10^{-8}$ [9]. We achieved a block-averaged secret key rate of 38.3 Mb/s and minimal secret key rate of 27.7 Mb/s over a 9.5 km fiber link using PCS 1024 QAM modulation, taking into account finite-size effects with a security parameter $\epsilon = 10^{-8}$. There are clear avenues for improvement, by moving to PCS 4096 QAM, and by increasing the block size. This result contributes in improving the performance of next generation high bit rate CV-QKD setups currently under development worldwide.

*Acknowledgement: This work has received funding from the European Union's Horizon 2020 research and innovation under grant agreements No 820466 CiViQ and No 857156 OpenQKD.*